\documentclass[showpacs,preprintnumbers,amsmath,amssymb,aps,twocolumn,superscriptaddress]{revtex4}

\usepackage{amsfonts}
\usepackage[dvips]{graphicx}
\usepackage{type1cm}
\usepackage{color}
\usepackage{bm}

\newcommand{\figref}[1]{Fig.~\ref{#1}}
\newcommand{\e}[1]{\text{e}^{#1}}
\newcommand{\cmplxi}{\text{i}}
\newcommand{\tr}{\operatorname{Tr}}
\renewcommand{\vec}[1]{\mathbf{#1}}
\newcommand{\punc}[1]{\,#1}
\newcommand{\neweqnline}{\nonumber\\}

\begin{document}
\title{Inhomogeneous phase formation on the border of itinerant ferromagnetism}
\author{G.J.~Conduit}
\email{gjc29@cam.ac.uk}
\affiliation{Cavendish Laboratory, 19, J.J.~Thomson Avenue, Cambridge, CB3 0HE. UK}
\author{A.G.~Green}
\affiliation{School of Physics and Astronomy, North Haugh, St Andrews, Fife, KY16 9SS. UK}
\author{B.D.~Simons}
\affiliation{Cavendish Laboratory, 19, J.J.~Thomson Avenue, Cambridge, CB3
  0HE. UK}

\date{\today}

\begin{abstract}
A variety of analytical techniques suggest that quantum fluctuations lead to a
fundamental instability of the Fermi liquid that drives ferromagnetic
transitions first order at low temperatures. We present both analytical and
numerical evidence that, driven by the same quantum fluctuations, this first
order transition is pre-empted by the formation of an inhomogeneous magnetic
phase. This occurs in a manner that is closely analogous to the formation of the
inhomogeneous superconducting Fulde-Ferrel-Larkin-Ovchinnikov state. We derive
these results from a field theoretical approach supplemented with numerical
Quantum Monte Carlo simulations. Our analytical approach represents a
considerable simplification over diagrammatic methods, makes contact with older
analyses of the unitarity limit, and enables a simple physical picture to
emerge.
\end{abstract}


\maketitle


Many magnetic materials display second order ferromagnetic phase transitions.
The transition temperature can be tuned using external parameters such as doping
and pressure. Hertz realized that tuning such a transition to zero temperature
could give rise to a new type of critical phenomena, for which he coined the
term quantum criticality~\cite{Hertz}. This led to a tremendous experimental and
theoretical effort that has had some notable successes with the universal
scaling predicted for the quantum critical regime being seen in a handful of
materials~\cite{Piers05}. However, in the majority of systems currently
investigated, new behavior intervenes before the transition temperature can be
tuned to zero. In many cases, such as $\text{ZrZn}_{2}$~\cite{Uhlarz04},
$\text{UGe}_{2}$~\cite{Huxley00}, MnSi~\cite{Pfleiderer01}, and
$\text{CoS}_{2}$~\cite{OteroLeal08}, the second order transition becomes first
order before the quantum critical point is reached.
Moreover, recent experimental evidence points to phenomena that go beyond the
first order transition; materials such as $\text{ZrZn}_{2}$~\cite{Uhlarz04},
$\text{UGe}_{2}$~\cite{Huxley01},
$\text{Ca}_{3}\text{Ru}_{2}\text{O}_{7}$~\cite{Baumberger06},
$\text{NbFe}_{2}$~\cite{Crook95}, and
$\text{Sr}_{3}\text{Ru}_{2}\text{O}_{7}$~\cite{Borzi07} all display unusual
behavior in the vicinity of the putative quantum critical point.

This failure to find a naked quantum critical point has lead to speculation that
it represents a fundamental principle~\cite{QuantumConundrum}.  Diagrammatic
calculations that extend beyond the standard Moriya-Hertz-Millis~\cite{Hertz}
theory of itinerant quantum criticality suggest a breakdown of the Landau
expansion around the quantum critical point~\cite{Shimzu64,Rech06,
BelitzReview05}.  This raises the question of how to connect the diagrammatic
calculations to the well-established second order perturbation
approach~\cite{Abrikosov58}. This approach accounts for all orders of vacuum
scattering amplitude, and predicts that the itinerant ferromagnetic transition
is first order at low temperature.

We show that, when a linearization of the electron dispersion about the Fermi
surface is permissible, the first order magnetic transition is pre-empted by the
formation of an inhomogeneous magnetic phase in a manner closely analogous to
the inhomogeneous superconducting Fulde-Ferrel-Larkin-Ovchinnikov (FFLO)
state~\cite{Fulde}. Our results are consistent with the effects of
non-analyticities that appear in extensions to the Hertz-Millis
theory~\cite{Shimzu64,Rech06}. Here, we adopt an alternative field theoretical
approach~\cite{Conduit08} to provide analytical evidence for
our picture. As well as resolving the connection between the second order
perturbation theory approach and the diagrammatic analysis, it provides insight
into how quantum fluctuations drive the reconstruction of the phase diagram. Our
analytical considerations are supported by Quantum Monte Carlo (QMC)
simulations.

This letter is divided into two main parts: We first develop our analytical
results and then discuss the QMC simulations that support them. Since one of the
main advantages of our approach is the simplification that it affords, we
outline the main steps of our analysis in their entirety. We follow this with a
discussion of the effect of quantum fluctuations upon the phase diagram of the
homogeneous ferromagnet -- showing both how older results may be recovered from
our analysis and the non-analyticities revealed by diagrammatics. Next, we show
how the same fluctuations can drive a spatial modulation. After presenting our
QMC results, we conclude with a discussion of possible extensions to the work.

In metals, the long-range component of the Coulomb interaction is screened. 
We therefore take as our starting point a free electron system interacting
through a contact (Hubbard-like) repulsive interaction, $g$~\cite{Hertz}.
The corresponding partition function may be expressed as a fermionic coherent 
state path integral $\mathcal{Z}=\tr\e{-\beta(\hat{H}-\mu \hat{N})}=
\int \mathcal{D}\psi\,\e{-S}$ with the action 
\begin{equation}
S=\int\sum_{\sigma=\pm} \bar\psi_\sigma \left(\partial_\tau
+\hat\zeta\right)\psi_\sigma
+\int g\bar\psi_+\bar\psi_-\psi_-\psi_+ \punc{.}
\label{stoneraction}
\end{equation}
Here $\int\equiv \int_0^\beta d\tau\int d^3x$, $\hat{\zeta}_{\hat{\bf k}}=
\epsilon_{\hat{\bf k}}-\mu$, where $\epsilon_{\hat{\bf k}}$ denotes the
dispersion and $\mu$ represents the chemical potential. To develop an effective
Landau theory of the magnetic transition, Hertz introduced a scalar
Hubbard-Stratonovich decoupling of the interaction in the spin
channel~\cite{Hertz}. However, this form of decoupling neglects the potential
impact of soft transverse field fluctuations.  It is the latter that are
responsible for driving the second order transition first order and in turn
promote the instability towards inhomogeneous phase formation. Therefore, we
will introduce a general Hubbard-Stratonovich decoupling which incorporates
fluctuations in all of the spin $\bm \Phi$ and charge $\rho$ sectors.

Defining $\bm \Phi={\bf m}+\bm \phi$, where ${\bf m}$ denotes the putative
saddle-point value of the magnetization field and $\bm \phi$ the fluctuations,
one obtains the partition function $\mathcal{Z}=\int\mathcal{D}\psi
\mathcal{D}\rho\mathcal{D}{\bm \phi}\e{-S}$ where the action now takes the form
$S=\int g(m^2+\phi^2-\rho^2)+\bar{\psi}[\partial_\tau+
\hat{\zeta}+g\rho-g{\bm\sigma}\cdot({\bf m}+{\bm\phi})]
\psi$~\cite{FootnoteCrossFluctuations}. We wish to study the potential for
spatially modulated order where the magnetization forms a conical spiral,
rotating about some axis (set along $z$) with a pitch vector ${\bf q}$. To
simplify the analysis, it is helpful to transform to a rotating basis in which
the magnetization becomes spatially uniform, ${\bf
m}=(m_\perp,0,m_\parallel)$. Setting $\psi\mapsto \e{\cmplxi{\bf q}\cdot{\bf
r}\,\sigma_{\text z}/2}\psi$,
\begin{eqnarray*}
\left(\begin{array}{c}\phi_{\text{x}}\\\phi_{\text{y}}\\\phi_{\text{z}}\end{array}\right)
\mapsto \left(\begin{array}{rrr}\cos(\vec{q}\cdot\vec{r})&-\sin(\vec{q}
\cdot\vec{r})&0\\
\sin(\vec{q}\cdot\vec{r})&\cos(\vec{q}\cdot\vec{r})&0\\
0&0&1\end{array}\right)
\left(\begin{array}{c}\phi_{\text{x}}\\\phi_{\text{y}}\\\phi_{\text{z}}\end{array}\right)\punc{,}
\end{eqnarray*}
and integrating over the fermion degrees of freedom yields the action
\begin{eqnarray}
&&S=\int g(m^2+\phi^2-\rho^2)\neweqnline
&&-\tr\ln[\underbrace{\partial_\tau+\hat\zeta_{\vec{k}+
\sigma_{\text{z}}\vec{q}/2}-g{\bm\sigma}\cdot\vec{m}}_{\hat{G}^{-1}}+g\rho
-g{\bm\sigma}\cdot{\bm\phi}]\punc{.}
\end{eqnarray}
Then, integrating over $\rho$ and $\phi$, an expansion of the action to 
second order in $g$ leads to 
\begin{widetext}
\vspace{-16pt}
\begin{eqnarray*}
{\cal Z}
=\exp\left[-\int g m^2+\tr\ln\hat{G}^{-1}-\tr g\hat{\Pi}^{+-}-
\frac{1}{2}\tr g^{2}(\hat{\Pi}^{+-}\hat{\Pi}^{-+}-\hat{\Pi}^{++}\hat{\Pi}^{--})\right]\punc{,}
\end{eqnarray*}
where $\hat{\Pi}^{ss'}=\hat{G}^s\hat{G}^{s'}$ denotes the Lindhard function and
$\hat{G}^\pm=(\partial_{\tau}+\epsilon_{\hat{\bf k},{\bf q}}^\pm-\mu)^{-1}$ with
\begin{equation}
\epsilon^{\pm}_{{\bf k},{\bf q}}=\frac{\epsilon_{{\bf k}+{\bf q}/2}+
\epsilon_{{\bf k}-{\bf q}/2}}{2}\pm\frac{\sqrt{(\epsilon_{{\bf k}-{\bf q}/2}-
\epsilon_{{\bf k}-{\bf q}/2}+2gm_\perp)^2+(2gm_\parallel)^2}}{2}
\end{equation}
\end{widetext}
representing the energy of the electrons in plane-wave states with momentum
${\bf k}$ and spin-up or down relative to the mean-field spiral. To remove the
unphysical ultraviolet divergence due to the contact nature of the interaction 
potential and arising from the term in the action second order in $g$, we must
affect the standard regularization of the linear term $\tr g\hat{\Pi}^{+-}$ 
setting $g\mapsto2k_{\text{F}}a/\pi\nu-2(2k_{\text{F}}a/\pi\nu)^{2}/V
\sum_{\vec{k}_{3,4}}'n(\epsilon^+_{{\bf k}_3,\vec{q}})n(\epsilon^-_{{\bf k}_4,
\vec{q}})(\epsilon^+_{{\bf k}_1,\vec{q}}+\epsilon^-_{{\bf k}_2,\vec{q}}-
\epsilon^+_{{\bf k}_3,\vec{q}}-\epsilon^-_{{\bf k}_4,\vec{q}}
)^{-1}$~\cite{Pathria07}, where the prime indicates that the summation is 
subject to the momentum conservation ${\bf k}_1+{\bf k}_2={\bf k}_3+{\bf
k}_4$. This regularization allows us to characterize the strength of the
interaction through the dimensionless parameter $k_{\text{F}}a$, where
$k_{\text{F}}$ denotes the Fermi wave vector, $a$ is the s-wave scattering 
length, and $\nu$ is the density of states at the Fermi surface.

Finally, carrying out the Matsubara summations, one obtains the following
expression for the free energy:
\begin{eqnarray}
&&F=\sum_{{\bf k},s}\epsilon_{{\bf k},{\bf q}}^{s}n\left(
\epsilon_{{\bf k},{\bf
q}}^{s}\right)+\frac{2k_{\text{F}}a}{\pi\nu V}N_{\vec{q}}^{+}N_{\vec{q}}^{-}\nonumber\\
 &&-
2\left(\frac{2k_{\text{F}}a}{\pi\nu V}\right)^{2}\sum_{\bf k}\int d\epsilon^+d\epsilon^-\frac{
\rho^+_{\vec{q}} ({\bf k}, \epsilon^+)\rho^-_{\vec{q}}(-{\bf k}, \epsilon^-)
}{\epsilon^++\epsilon^-}
\nonumber\\
&&+2\left(\frac{2k_{\text{F}}a}{\pi\nu V}\right)^{2}\!\!{\sum_{{\bf k}_{1,2,3,4}}\!\!}'
\frac{n(\epsilon_{{\bf k}_1,\vec{q}}^{+})n(\epsilon_{{\bf k}_2,\vec{q}}^{-})}
{\epsilon^+_{{\bf k}_1,\vec{q}}+\epsilon^-_{{\bf k}_2,\vec{q}}-
\epsilon^+_{{\bf k}_3,\vec{q}}-\epsilon^-_{{\bf k}_4,\vec{q}}}\punc{,}
\label{FreeEnergy}
\end{eqnarray}
where $n(\epsilon)=1/(1+\e{\beta(\epsilon-\mu)})$ 
is 
the Fermi distribution,
$N^{s}_{\vec{q}}=\sum_{\vec{k}}n(\epsilon_{\vec{k},\vec{q}}^{s})$, and
$\rho^\pm_{\vec{q}}({\bf k},\epsilon)=\sum_{\vec{k}'}
n(\epsilon_{\vec{k}'+\vec{k}/2,\vec{q}}^{\pm})[1-n(\epsilon_{\vec{k}'-\vec{k}/2,\vec{q}}^{\pm})]
\delta(\epsilon-\epsilon_{\vec{k}'+\vec{k}/2,\vec{q}}^\pm+\epsilon_{\vec{k}'-\vec{k}/2,\vec{q}}^\pm)$ is
the spin-up(down) particle-hole density of states.

\begin{figure}
\centerline{\resizebox{0.95\linewidth}{!}{\includegraphics{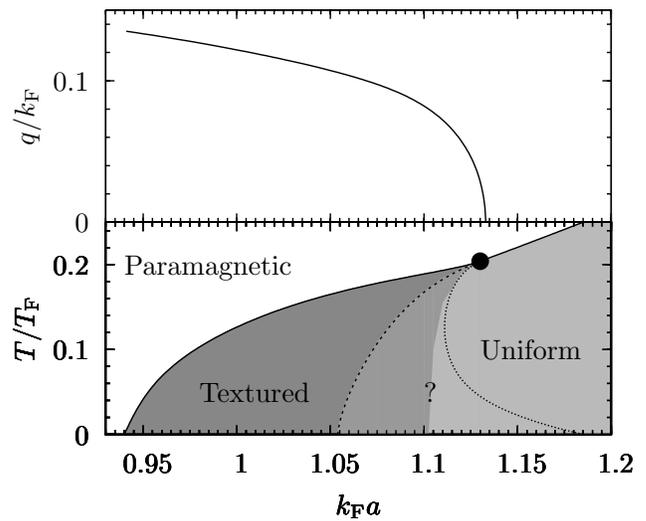}}}
 \caption{The lower phase diagram shows the first order (dashed line) and second
   order (upper solid line) ferromagnetic ordering if the system were restricted
   to be homogeneous. The first order transition is pre-empted by a second order
   phase transition into a modulated ferromagnetic state bounded by the solid
   line. The dotted line shows where the uniform ferromagnetic phase would form,
   if the transition were restricted to be second order; this putative phase
   boundary not only comes after the first order transition and joins smoothly
   to the second order phase boundary at temperatures above the tricritical
   point, but also has negative slope at $T=0$, consistent with
   Refs.~\cite{Rech06}. The upper graph shows the wave vector $q$ of the
   modulated phase boundary.}
 \label{fig:ModItinFerro}
\end{figure}

{\it Spatially Uniform Case:} To orient our discussion, we first consider the
implications of the fluctuation corrections on the magnetic phase diagram for a
free particle dispersion $\epsilon_{\vec{k}}=k^{2}/2m^{*}$ without accounting
for spatial modulation. In this case Eq.~(\ref{FreeEnergy}) reduces to that
obtained in Ref.~\cite{Abrikosov58} from second order perturbation theory. From
this result, one finds that fluctuations drive the second order ferromagnetic
transition first order at temperatures below that of the tricritical point,
$T_{\rm T}\approx 0.2T_{\text{F}}$, where $T_{\text{F}}$ is the Fermi
temperature (see \figref{fig:ModItinFerro}).  Substituting the low energy form
of the particle-hole density of states, $\rho^\pm_{{\bf q}={\bf 0}}({\bf k},
\epsilon)= \epsilon \theta(k k_{\text{F}}^\pm-k^2/2-\epsilon)/2 \pi k$,
where $k_{\text{F}}^\pm=\sqrt{2m^*(\mu\pm 2k_{\text{F}}am/\pi\nu)}$, into the 
fluctuation correction to the free energy, and expanding in powers of
magnetization, one recovers a singular term of order $m^4 \ln m^2$ arising from
particle-hole excitations with momentum near to $2k_{\text F}$, the same
non-analyticity as was found diagrammatically in
Refs.~\cite{Shimzu64,Rech06}. The formation of a finite magnetization increases
the phase-space available for the formation of virtual intermediate pairs of
particle-hole pairs, and this phase space enhancement ultimately drives the
ferromagnetic transition first order.


{\it Spatially Modulated Case:} Quantum fluctuations also lead to spatial
modulation of the magnetic order and further reconstruction of the magnetic
phase diagram. The results of this analysis are shown in
\figref{fig:ModItinFerro}, which is obtained from Eq.~(\ref{FreeEnergy}) through
a Landau expansion in $m$~\cite{FootnoteLandauExpansion}.  At low temperatures
and $k_{\text{F}}a=0.84$, a second order transition into an inhomogeneous spin
phase with pitch $0.1k_{\text{F}}$ pre-empts the first order phase transition at
$k_{\text{F}}a=1.055$. The expansion cannot however describe how far the
textured phase penetrates into the uniform ferromagnetic phase. For a quadratic
dispersion, the magnetization and spiral wave-vector enter the single electron
energy in the combination $({\bf k}\cdot{\bf
q}/m^*)^2+(4k_{\text{F}}am_\perp/\pi\nu)^2$, where we have
restricted to a planar spiral since it always has lower energy than a conical
spiral at zero magnetic field. Upon linearizing the electron energy at the Fermi
surface, $q^2$ and $m_\perp^2$ enter the free energy in the same way (up to the
angular factors that accompany ${\bf q}$ and which are integrated
over). Therefore, the spatial modulation enters as if it where a direction
dependent magnetization.  As a consequence, the coefficient of $m_\perp^4$ in
our Taylor expansion is proportional to the coefficient of $q^2 m_\perp^2$. When
the $m_\perp^4$-term becomes negative -- and a first order transition becomes
favorable -- the $q^2 m_\perp^2$-term also becomes negative favoring a spatial
modulation that emerges from the tricritical point. This is precisely the same
situation as seen in the FFLO state~\cite{Fulde}.

To assess the validity of the perturbative scheme, we turn now to the 
numerical Quantum Monte Carlo analysis of the Stoner 
Hamiltonian~(\ref{stoneraction}) making use of the CASINO 
program~\cite{Needs08}. These methods are based upon optimizing a 
trial wave function and are restricted to zero temperature. Our approach 
mirrors that used in previous studies of itinerant 
ferromagnetism~\cite{Ceperley80,Ortiz99,Zong02}. The variational wave 
function used in our simulation, $\psi=D\e{-J}$, is a product of a 
Slater determinant, $D$, that takes account of the Fermion statistics 
and occupation of single particle orbitals, and a Jastrow factor, $J$, 
that accounts for electron correlations.

{\it The Slater determinant} consists of plane-wave spinor orbitals containing
both spin-up and spin-down electrons, $D=\det(\{\psi_{\vec{k}\in k_{\uparrow}},
\bar{\psi}_{\vec{k}\in k_{\downarrow}}\})$. Although not an eigenstate of total
spin, the constituent spin states of least weight provide the dominant
contribution to the variational state energy. In the case of uniform
magnetization, for computational efficiency, we factorize the Slater determinant
into an up and a down-spin determinant~\cite{Needs08}. The spin textured phase
is described by non-collinear spins, which have only recently been studied
within the Variational Monte Carlo (VMC)~\cite{Radnai}. These studies lead us to
describe a planar spin spiral with a trial wave function that contains the
spinors
$\psi_{\vec{k}}=\e{\cmplxi\vec{q}\cdot\vec{r}/2}(\e{\cmplxi\vec{k}\cdot\vec{r}},\e{-\cmplxi\vec{k}\cdot\vec{r}})$
and
$\bar{\psi}_{\vec{k}}=\e{-\cmplxi\vec{q}\cdot\vec{r}/2}(-\e{\cmplxi\vec{k}\cdot\vec{r}},\e{-\cmplxi\vec{k}\cdot\vec{r}})$
which explicitly fix the spin spiral orientation. For $\vec{q}=\vec{0}$ this
would recover the factorized form for the Slater determinant employed in the
uniform case. Our simulations are carried out in a unit cell with periodic
boundary conditions commensurate with the pitch of the spiral.

\begin{figure}
 \centerline{\resizebox{0.95\linewidth}{!}{\includegraphics{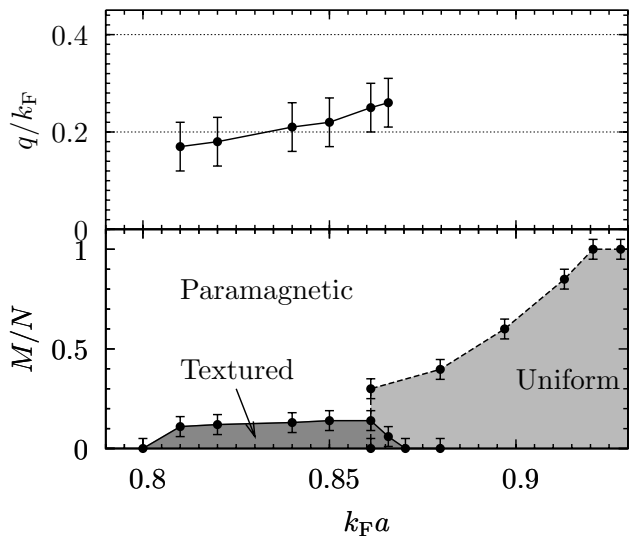}}}
 \caption{The lower panel shows the variation of the ground state 
   magnetization $M$ with interaction strength $k_{\text{F}}a$ at zero 
   temperature. The dashed line corresponds to the uniform phase, and the 
   solid line is the textured phase. The upper panel shows the wave vector 
   of the inhomogeneous magnetic phase, and the discrete values of $q$ 
   sampled in the investigation are highlighted by the horizontal dotted 
   lines.}
 \label{fig:CASINOFerroPhaseDiag}
\end{figure}

{\it The Jastrow factor}, $J$, accounts for electron-electron correlations. It
consists of polynomial and plane-wave expansions in electron-electron
separation, and a polynomial backflow function. In the spiral case, the Jastrow
factor is restricted to be spin independent to maintain the spin spiral
orientation and the wave function antisymmetry. The trial wave functions were
optimized using QMC methods. In the uniform case, the optimization was performed
in two steps using VMC and Diffusion Monte Carlo, whereas only VMC calculations
were performed for the textured state. To model the repulsive contact potential
between the electrons we employ the modified P\"oschl-Teller
interaction~\cite{Carlson03} which has smooth edges so that the QMC
configurations can sample it faithfully~\cite{FootnoteQMCTesting}.

Firstly, constraining the magnetization to be spatially uniform, an estimate of
the ground state magnetization for different interaction strengths recovers the
expected phase diagram (\figref{fig:CASINOFerroPhaseDiag}), revealing a first
order phase transition into the itinerant magnetic phase.  This provides a
platform upon which to construct the full textured phase diagram. Allowing for
spatial modulation of the magnetization, we find that an inhomogeneous magnetic
phase pre-empts the transition into the uniform phase. The resulting textured
phase has similar extent and wave vector to the analytical prediction lending
support to the conclusions of the perturbative field theoretic analysis.

In conclusion, quantum fluctuations are known to drive the itinerant
ferromagnetic transition first order. We have shown that the same mechanisms
lead to the development of inhomogeneous magnetic order in the vicinity of the
tricritical point, resulting in a phase diagram which mirrors that of the
superconducting FFLO phase. Our results are consistent with recent diagrammatic
analyses and reveal the connection between these works~\cite{Shimzu64,Rech06}
and older second order perturbation calculations~\cite{Abrikosov58}. Moreover,
our approach represents a considerable analytical simplification and more
clearly reveals the underlying physical processes.

There are several directions in which this analysis might be extended. An
electronic band dispersion can drive a similar reconstruction of the electronic
phase diagram~\cite{Berridge08}. It would be informative to investigate the
interplay between such effects and quantum fluctuations. As well as the
possibility of spatially modulated magnetism, recent works have suggested 
that a band dispersion might lead to d-wave distortion of the Fermi
surface~\cite{Kee05}. Such electron nematics may be viewed as melted
versions of the spatially modulated magnetism that we consider. Indeed, 
quantum fluctuations might drive the formation of an electron nematic phase 
in itinerant ferromagnets.

{\it Acknowledgments:} We thank Andrew Berridge, Joseph Betouras, Andrei
Chubukov, Una Karahasanovic, and Andrew Schofield for useful discussions. Neil
Drummond and Zoltan Radnai provided invaluable assistance with the QMC
calculations. The authors acknowledge the financial support of the EPSRC, the
Royal Society, and the Miller Institute.


\begin{thebibliography}{99}
\bibitem{FootnoteLandauExpansion}
The focus on non-zero temperatures leads to
a regularization of the various non-analyticities in the magnetization and its
gradients, legitimizing the Landau expansion.
\bibitem{FootnoteQMCTesting}
To ensure that the well was a
suitable model for the contact interaction we checked that the results did not
depend on different well parameters that have the same scattering length, and
that a square well potential yields the same ground state energy. Results were
extrapolated to zero time step, and twist averaging was employed to compensate
for the finite system size.
\end{thebibliography}
\end{document}